\documentclass[10pt]{amsart}
\usepackage{amsmath}
\usepackage{latexsym}
\usepackage{amsfonts}
\usepackage{amssymb}
\usepackage{color}
\usepackage{bbm,dsfont}
\usepackage{amsthm}

\newtheorem{theorem}{Theorem}

\newcommand{\R}{\mathbb R} 

\newcommand{\hi}{\mathcal H} 
\newcommand{\lh}{\mathcal{L(H)}} 
\newcommand{\trh}{\mathcal{T(H)}} 
\newcommand{\ip}[2]{\langle {#1}|{#2}\rangle} 
\newcommand{\tr}[1]{\mathrm{tr}\left[ {#1} \right]} 
\newcommand{\kb}[2]{|#1\,\rangle\langle\,#2|} 

\newcommand{\stfm}{{\mathcal I}} 
\newcommand{\ltwo}{L^2(\R)}

\newcommand{\sfq}{\mathsf{Q}}
\newcommand{\sfp}{\mathsf{P}}
\newcommand{\qhat}{\hat{Q}}
\newcommand{\phat}{\hat{P}}

\newcommand{\br}{\mathcal{B}(\R)} 
\newcommand{\prob}{\mathsf{p}} 

\newcommand{\N}{\mathcal{N}}

\newcommand{\veps}{\varepsilon}
\newcommand{\eps}{\epsilon}

\newcommand{\dele}[1]{\mathcal{W}_{\varepsilon,\delta}({#1},E)}

\newcommand{\deq}[1]{\mathcal{W}_{\varepsilon_1}({#1},\sfq)}
\newcommand{\dee}[1]{\mathcal{W}_{\varepsilon}({#1},E)}
\newcommand{\dep}[1]{\mathcal{W}_{\varepsilon_2}({#1},\sfp)}

\begin{document}
\title[Heisenberg's Uncertainty Principle]{Heisenberg's Uncertainty Principle}

\author[Busch]{Paul Busch}
\author[Heinonen]{Teiko Heinonen}
\author[Lahti]{Pekka Lahti}

\begin{abstract}
\noindent Heisenberg's
uncertainty principle is usually taken to express a \emph{limitation} of
operational possibilities imposed by quantum mechanics. Here we demonstrate
that the full content of this principle also includes its \emph{positive}
role as a condition ensuring that mutually exclusive experimental options can 
be reconciled if an appropriate trade-off is accepted. The uncertainty principle is 
shown to appear in three manifestations, in the form of uncertainty relations:
for the widths of the position and momentum distributions in
any quantum state; for the inaccuracies of any joint measurement of these
quantities; and for the inaccuracy of a measurement  of one of the quantities
and the ensuing disturbance in the distribution of the other quantity. Whilst
conceptually distinct, these three kinds of uncertainty relations
are shown to be closely related formally. 
Finally, we survey models and experimental
implementations of joint measurements of position and momentum and comment
briefly on the status of experimental tests of the uncertainty principle.
\end{abstract}

\thanks{This work has been published in \emph{Physics Reports} 452 (2007) 155-176}

\maketitle

\vspace{12pt}

\begin{quote}
``It is the theory which decides what can be observed."\\
{\small (Albert Einstein according to Werner Heisenberg \cite{Heisenberg77})} 
\end{quote}

\section{Introduction}\label{intro}

It seems to be no exaggeration to say that Heisenberg's uncertainty principle, symbolized by the 
famous inequality for position and momentum,
\begin{equation}\label{qp-ur}
\Delta q\cdot\Delta p\gtrsim\hbar,
\end{equation}
epitomizes quantum physics, even in the eyes of the scientifically informed public. 
Nevertheless, still now, 80 years after its inception, there is no general consensus 
over the scope and validity of this principle.  The aim of this article is to demonstrate
that recent work has finally made it possible to elucidate the full content 
of the uncertainty principle in precise terms. This will be done for the prime example of
the position-momentum pair.

The uncertainty principle is usually described, rather vaguely, as
comprising one or more of the following \emph {no-go} statements,
each of which will be made precise below:
\begin{enumerate}
\item[(A)] \emph{It is impossible to prepare states in which position and momentum
are simultaneously arbitrarily well localized.}
\item[(B)] \emph{It is impossible to measure simultaneously position and momentum.}
\item[(C)] \emph{It is impossible to measure position without disturbing momentum, and vice versa.}
\end{enumerate}

The negative characterization of the uncertainty principle as a limitations of
quantum preparations and measurements has led to the widespread view
that this principle is nothing but a formal expression of the principle of 
complementarity.\footnote{A more balanced account of the interplay and relative 
significance of the principles of complementarity and uncertainty has been developed 
in a recent review \cite{BuSh06} which complements the present work.}
This limited perspective has led some authors to question the fundamental status
of the relation (\ref{qp-ur}) \cite{ScEnWa91}. 

Here we will show that the uncertainty principle does have an independent content and role 
that in our view has not yet been duly recognized. In fact, instead of resigning to accept the 
negative verdicts (A), (B), (C), it is possible to adopt a
positive perspective on the underlying questions of joint
preparation and measurement: according to the uncertainty principle,
the qualitative relationship of a strict mutual exclusiveness of sharp preparations 
or measurements of position and momentum is
complemented with a quantitative statement of a trade-off between
competing degrees of the concentration of the distributions of these observables in state
preparations or between the accuracies in joint measurements. Similarly, it turns out 
that the extent of the disturbance of (say) momentum in a measurement of position can 
be controlled if a limitation in the accuracy of that measurement is accepted.

Only if taken together, the statements (A), (B), (C) and their positive counterparts
can be said to exhaust the content of
the uncertainty principle for position and momentum. It also follows that the
uncertainty principle comprises three conceptually distinct types of
uncertainty relations. 

We will give a systematic exposition of these three faces of the uncertainty principle, with an
emphasis on elucidating its positive role. After a brief discussion of the well known uncertainty
relation for preparations, we focus on the less well established \emph{measurement} uncertainty
relations, the formulation of which requires a careful discussion of joint measurements, measurement accuracy and disturbance.

We present a fundamental result, proved only
very recently, which constitutes the first rigorous demonstration
of the fact that the uncertainty relation for measurement inaccuracies is not only a sufficient
but also a necessary condition for the
possibility of approximate joint measurements of position and momentum.

Finally, we discuss  some models
and proposed realizations of joint measurements of position and momentum and
address the question of possible experimental tests
of the uncertainty principle.

The idea of the uncertainty principle ensuring the
positive possibility of joint albeit imprecise measurements, which is rather
latent in Heisenberg's works\footnote{A judicious reading of Heisenberg's seminal paper of 1927 \cite{Heisenberg27} shows that both the double role
and the three variants of the uncertainty principle discussed here are already manifest, if only
expressed rather vaguely. In fact, in the abstract, Heisenberg immediately refers to limitations
of joint measurements; later in the paper, he links this with a statement of the uncertainty relation
for the widths of a Gaussian wave function and its Fourier transform; finally he gives illustrations
by means of thought experiments in which the idea of mutual disturbance is prominent.}
has been made fully explicit and brought to our
attention by his former student Peter Mittelstaedt, our teacher and mentor, to
whom we dedicate this treatise.

\section{From ``no joint sharp values" to  approximate joint localizations}\label{prep-ur}

Throughout the paper, we will only consider the case of a spin zero quantum
system in one spatial dimension, represented by the Hilbert space $\hi=\ltwo$.
The states of the system are described by positive trace
one operators  $\rho$ on $\hi$, 
the pure states being given as the one-dimensional projections.
We occasionally write $\kb\psi\psi$ for the pure state in question, and we call unit vectors 
$\psi\in\hi$ \emph{vector states}. We denote the set of vector states by $\hi_1$.

The position and momentum of the system are represented as the
Schr\"odinger pair of operators $\qhat,\phat$, where $\qhat\psi(x)=x\psi(x)$,
$\phat\psi(x)=-i\hbar\psi'(x)$. We denote their spectral measures by the
letters $\sfq$ and $\sfp$, respectively, and recall that they are
Fourier-Plancherel connected. The probability of obtaining the value of
position in a (Borel) subset $X$ of $\mathbb{R}$ on measurement in a
vector state $\psi$ is then given by the formula 
$\prob_\psi^{\sfq}(X)=
\ip{\psi}{\sfq(X)\psi} =\int_X |\psi(x)|^2\,dx$.
Similarly, the probability of obtaining the value of
momentum in a (Borel) set $Y$ on measurement in a vector state $\psi$ is given by
$\prob_\psi^{\sfp}(Y) := \int_Y|\widehat{\psi}(p)|^2\,dp$, where
$\widehat{\psi}$ is the Fourier-Plancherel transform of $\psi$. 

In the formalization of all three types of uncertainty relations, we will use two different
measures of the width, or degree of concentration, of a probability distribution. These are
the standard deviation and the overall width. The \emph{standard
deviations} of position and momentum in a state $\psi$ are 
\begin{equation}\begin{split}
\Delta(\sfq,\psi)&:=\left(\ip{\psi}{\qhat^2\psi}-\ip{\psi}{\qhat\psi}^2\right)^{1/2},\\
\Delta(\sfp,\psi)&:=\left(\ip{\psi}{\phat^2\psi}-\ip{\psi}{\phat\psi}^2\right)^{1/2}.
\end{split}
\end{equation}

The \emph{overall width} of a probability measure $\prob$ on $\R$ is defined, for given  
$\varepsilon\in(0,1)$, as the smallest interval length required to have probability greater or 
equal to $1-\varepsilon$; thus:
\begin{equation}
W_\veps(\prob):=\inf_X\{|X|\,|\,\prob(X)\ge 1-\varepsilon\},
\end{equation}
where $X$ run through all intervals in $\R$.
For the overall widths of the position and momentum distribution in a vector state $\psi$ we will 
use the notation
\begin{equation}\label{overall-width}
W_{\veps_1}(\sfq,\psi):=W_{\veps_1}(\prob_\psi^\sfq),
\qquad
W_{\veps_2}(\sfp,\psi):=W_{\veps_2}(\prob_\psi^\sfp).
\end{equation}

The no-go statement (A) says, broadly speaking, that the distributions $\prob^\sfq_\psi$ and
$\prob^\sfp_\psi$ of position and momentum
cannot simultaneously (i.e., in the same state $\psi$) be arbitrarily sharply concentrated.  
To appreciate this fact,
we note first that position and momentum, being continuous quantities, cannot
be assigned absolutely sharp values since they have no eigenvalues.
But both quantities can separately have arbitrarily sharply concentrated distributions. 
We discuss two different ways of formalizing this idea, in terms of standard deviations and 
overall widths. Each of these formalizations gives rise to 
a precise form of (A).  We then proceed to complement the no-go statement (A) with descriptions 
of the positive possibilities of simultaneous \emph{approximate localizations} of position and
momentum.

The first formalization of arbitrarily sharp localizations makes use of the standard deviation 
of a distribution (stated here for position):
\begin{equation}\begin{split}
& \text{for any}\ q_0\in\R\ \text{and any\ }\varepsilon>0,\ \text{there is a vector state}\
\psi\\ 
&\text{such that\ } \ip{\psi}{\qhat\psi}=q_0\ \text{and} \ \Delta(\sfq,\psi)<\varepsilon.\label{sv2}
\end{split}\end{equation}
Thus there is no obstacle to concentrating the distributions of position or momentum 
arbitrarily sharply at any points $q_0,p_0\in\R$, if these observables are considered separately, 
on different sets of states.

But the corresponding state preparation procedures are mutually exclusive; this is the
operational content
of the negative statement (A). What \emph{is} positively possible if one considers both observables together in the same state will be described by an appropriate uncertainty relation. Property (\ref{sv2}) gives rise to the following formalization of (A):  

\begin{theorem}\label{no-go-1b}
For all states $\psi$ and for any $\varepsilon>0$,
\begin{equation}\label{neg-1b}
\textrm{if}\quad\Delta(\sfq,\psi)  <  \varepsilon,
\quad\textrm{then}\quad \Delta(\sfp,\psi)>\hbar/2\varepsilon; \quad\text{and vice versa}.
\end{equation}
\end{theorem}
This is a statement about the spreads of the position and
momentum probability distributions in a given state: the sharper one
is peaked, the wider the other must be. 
This limitation follows directly from the uncertainty relation for standard
deviations, valid for all vector states $\psi$:
\begin{equation}\label{p-ur}
\Delta(\sfq,\psi)\cdot\Delta(\sfp,\psi)\ge \frac\hbar 2.
\end{equation}

The vector states $\eta_{a,b}(x)= (2a/\pi)^{1/4}\,e^{-(a+ib)x^2}$, 
$a,b\in\R$, $a>0$, give
 $\Delta(\sfq,\eta_{a,b})^2=  {1}/{4a}$ and 
 $\Delta(\sfp,\eta_{a,b})^2= \hbar^2(a^2+b^2)/{a}$, so that
the following positive statement complementing the no-go Theorem \ref{no-go-1b} is obtained:
\begin{theorem}\label{urnumbers}
For all positive numbers $\delta q,\delta p$ for which $\delta
q\cdot\delta p\ge\hbar/2$, there is a state $\psi$ such that
$\Delta(\sfq,\psi)=\delta q$ and $\Delta(\sfp,\psi)=\delta p$.
\end{theorem}

The vector state $\eta_{a,0}$ is a \emph{minimal uncertainty state} in the sense that it gives 
$\Delta(\sfq,\eta_{a,0})\cdot\Delta(\sfp,\eta_{a,0})= \hbar/2$.
Every minimum uncertainty state is of the form $e^{icx}\eta_{a,0}(x-d)$ for some $c,d\in\R$. 
Minimal uncertainty states have a number of distinctive properties (see, e.g., \cite{FSI}). 
For instance, if $\psi$ is a vector state which satisfies $|\psi |^2\leq |\eta_{a,0}|^2$ and 
$|\widehat{\psi}|^2\leq |\widehat{\eta_{a,0}}|^2$, 
for some $a$, then $\psi$ is a minimal uncertainty state.

We now turn to the second way of saying that position and momentum can separately be
localized arbitrarily well (expressed again only for position):
\begin{equation}\begin{split}
&\text{for any bounded interval $X$ (however small), there exists a vector state} \ \psi \\
&\text{such that}\ \prob_\psi^{\sfq}(X)=1.\label{sv1}
\end{split}\end{equation}
The corresponding formalization of (A) then  is given by the following theorem.
\begin{theorem}\label{no-go-1a}
For all vector states $\psi$ and all bounded intervals $X,Y$,
$\prob_\psi^{\sfq}(X) = 1$ implies $0\ne\prob_\psi^{\sfp}(Y)\ne 1$, and vice versa.
\end{theorem}
This means that whenever the position is localized in
a bounded interval then the momentum cannot be confined to any bounded interval (nor to
its complement), and vice versa.

For any two bounded intervals $X$ and $Y$ and for any vector state $\psi$,
Theorem~\ref{no-go-1a} implies that $p_\psi^\sfq(X)+p_\psi^\sfp(Y)<2$. However,
for any such intervals, one can  construct a vector state $\psi_0$ for which
the sum of the probabilities $\prob_{\psi_0}^\sfq(X)$ and
$\prob_{\psi_0}^\sfp(Y)$ attains its maximum value. The precise statement is
given in the following theorem, which can be regarded as a positive complement
to Theorem \ref{no-go-1a}.

\begin{theorem}
For any vector state $\psi$ and for any bounded intervals $X$ and $Y$,
\begin{equation}\label{probsum}
\prob_\psi^{\sfq}(X)+\prob_\psi^{\sfp}(Y)\leq1+\sqrt{a_0}<2,
\end{equation}
where $a_0$ is the largest eigenvalue of the operator $\sfq(X)\sfp(Y)\sfq(X)$ which is
positive and trace class. There exists an optimizing vector state $\varphi_0$
such that
\begin{equation}\label{optimstate}
\prob_{\varphi_0}^{\sfq}(X)+\prob_{\varphi_0}^{\sfp}(Y)=1+\sqrt{a_0}.
\end{equation}
\end{theorem}
This result follows from the work of Landau and Pollak \cite{LaPo61} and
Lenard \cite{Lenard72} (for details, see \cite{Lahti86}).

We will say that position $\sfq$ is \emph{approximately localized} in an
interval $X$ for a given state $\psi$ whenever $\prob_\psi^\sfq(X)\ge 1-\varepsilon$
for some (preferably small) $\varepsilon$, $0<\varepsilon<1$, and similarly for
momentum. Then Eq.~(\ref{optimstate}) describes the maximum degree of
approximate localization that can be achieved in any phase space cell of given size
$|X|\cdot |Y|$.

The largest eigenvalue $a_0$ is invariant under a scale
transformation applied simultaneously to $\qhat$ and $\phat$; it is
therefore a function of the product $|X|\cdot|Y|$ of the interval lengths $|X|$
and $|Y|$. A simple calculation gives  $\tr{\sfq(X)\sfp(Y)\sfq(X)}=|X|\cdot |Y|/(2\pi\hbar)$,
so that we obtain
\begin{equation}\label{widthbound}
{|X|\cdot |Y|} \ge {2\pi\hbar}\cdot a_0.
\end{equation}
If position and momentum are both approximately 
localized within $X$ and $Y$, respectively, so that $\prob_\psi^\sfq(X)\ge 1-\varepsilon_1$ and
$\prob_\psi^\sfp(Y)\ge 1-\varepsilon_2$, then due to inequality (\ref{probsum}), one must have
$1-\varepsilon_1-\varepsilon_2\le\sqrt{a_0}$, and then (\ref{widthbound}) implies:
\begin{equation}
|X|\cdot |Y| \ge {2\pi\hbar}\cdot (1-\varepsilon_1-\varepsilon_2)^2
\end{equation}
if $\prob_\psi^\sfq(X)\ge 1-\varepsilon_1$,
$\prob_\psi^\sfp(Y)\ge 1-\varepsilon_2$, and  $\varepsilon_1+\varepsilon_2<1$.

It is convenient to express this uncertainty relation for approximate localization widths 
in terms of the overall widths: if $\varepsilon_1+\varepsilon_2<1$ then
\begin{equation}\label{ow-ineq}
{W_{\veps_1}(\sfq,\psi)\cdot W_{\veps_2}(\sfp,\psi)}\ge
{2\pi\hbar}\cdot\left( 1-\varepsilon_1-\varepsilon_2 \right)^2.
\end{equation}
If $\varepsilon_1+\varepsilon_2\ge 1$, then the product of widths has no positive lower bound
\cite{LaPo61}.
In the case $\varepsilon_1+\varepsilon_2<1$, the inequality is tight in the sense that even for
fairly small values of $\varepsilon_1,\varepsilon_2$, the product of overall widths can be in
the order of $2\pi\hbar$; we quote a numerical example given in \cite{LaPo61}: if $\varepsilon_1=
\varepsilon_2=.01$, then $|X|\cdot|Y|$ can still be as small as $6.25\times(2\pi\hbar)$.

We will make repeated use of this uncertainty relation; since we are only interested in ``small" values of 
$\varepsilon_1,\varepsilon_2$, we will assume these to be less than 1/2; then the condition
$\varepsilon_1+\varepsilon_2<1$ is fulfilled and need not be stated explicitly.

An inequality of the form (\ref{ow-ineq}) has been given by J.B.M. Uffink in his doctoral thesis of 1990 \cite{Uffink90}; using a somewhat more involved derivation, he obtained the  sharper lower bound $2\pi\hbar \cdot \left(\sqrt{(1-\varepsilon_1)(1-\varepsilon_2)}-\sqrt{\varepsilon_1\varepsilon_2}\right)^2$.

Several other measures of uncertainty have been introduced to
analyze the degree of (approximate) localizability of position and
momentum distributions $\prob_\psi^{\sfq}$ and $\prob_\psi^{\sfp}$, ranging from
extensive studies on the support properties of $|\psi|^2$ and
$|\widehat{\psi}|^2$ to various information theoretic (``entropic") uncertainty
relations. It is beyond the scope of this paper to review the
vast body of literature on this topic. The interested reader may consult e.g.
\cite{CoPr84}, 
\cite{FoSi97} or \cite[Sect. V.4]{OQP} for reviews and references.
 
To summarize: instead of leaving it at the negative statement that
position and momentum \emph{cannot be arbitrarily
sharply localized in the same state}, the uncertainty relation for state preparations
offers  precise specifications of
the extent to which these two  observables \emph{can simultaneously
be approximately localized}.

\section{Joint and sequential measurements}\label{jsm}

In order to go beyond the no-go theorems of (B) and (C) and establish their positive complements,
one needs to use the full-fledged apparatus of quantum mechanics. The general
representation of observables as positive operator measures will be required 
to introduce viable notions of joint and sequential measurements and an appropriate quantification of measurement inaccuracy. Furthermore, some tools of measurement theory will be needed to describe and quantify the disturbance of one observable due to the measurement of another.

In the present context of the discussion of position and momentum and their joint measurements,
observables will be described as normalized positive operator measures $X\mapsto E(X)$ on the (Borel) subsets of $\R$ or $\R^2$.
This means that the map $X\mapsto \ip{\psi }{E(X)\psi}=:\prob_\psi^E(X)$ is a probability measure for every vector state $\psi$. The operators $E(X)$ in the range of an observable $E$ are called \emph{effects}. An observable $E$ will be called \emph{sharp} if it is a spectral measure, that is, if all of the effects  $E(X)$ are projections.\footnote{For a more detailed technical discussion of the notion of a quantum observable as a positive operator measure, the reader may wish to consult, for example, the monograph \cite{QTOS}; a gentle, less formal, introduction may be found in the related review of Ref. \cite{BuSh06}.} 

For an observable $E$ on $\R$, we will make use of the notation  $E[1],\ E[2]$ for the first and the second moment operators, 
defined (weakly) as $E[k]:=\int x^kE(dx)$ ($k=1,2$).\footnote{It is important to bear in mind that the domain of the symmetric operator $E[k]$ is not necessarily dense; it
consists of all vectors $\varphi\in\hi$ for which the function $x\mapsto x^k$ is integrable w.r.t. the complex measure $X\mapsto \ip{\psi}{E(X)\varphi}$ for all $\psi\in\hi$. Here and in subsequent formulas it is understood that the expectations of unbounded operators are only well defined for appropriate
subsets of states.} We let
 $\Delta(E,\psi)$ denote the standard deviation of $\prob_\psi^E$,
\begin{equation}\begin{split}
\Delta(E,\psi)^2&:=\int_{-\infty}^\infty \left(x-\int_{-\infty}^\infty
x'\prob_\psi^E(dx')\right)^2\prob_\psi^E(dx) \\ &
\ =\ip{\psi}{ E[2]\psi}-\ip{\psi}{ E[1]\psi }^2.
\end{split}\end{equation}

It is a remarkable feature of an observable $E$, defined as a positive operator measure, 
that it need not be commutative; that is, it is not always the case that 
$E(X_1)E(X_2)=E(X_2)E(X_1)$ for all sets $X_1,X_2$. This opens up the possibility of 
defining a notion of joint measurability for not necessarily  commuting families of observables. 

Indeed, it will become evident below that in the set of noncommuting pairs of observables,
the jointly measurable ones are necessarily unsharp, that is, they cannot be sharp.
It is to be expected intuitively that the degree of mutual noncommutativity determines the
necessary degree of unsharpness required to allow a joint measurement. Here we present two
ways of  indicating the inherent unsharpness of an observable $E$ on $\R$.

We define the \emph{intrinsic noise operator} of $E$ as
\begin{equation}
N_i(E):=E[2]-E[1]^2.
\end{equation}
This is a positive operator.  If  $E[1]$ is selfadjoint, then
the intrinsic noise $N_i(E)$ is zero exactly when $E$ is a sharp observable \cite[Theorem 5]{KiLaYl06b}. 
A measure $\N_i(E;\psi)$ of  \emph{intrinsic noise} is then given by the expectation value of the 
intrinsic noise operator (for all vector states $\psi$ for which this expression is well defined):
\begin{equation}
\N_i(E;\psi):=\ip{\psi}{N_i(E)\psi}.
\end{equation}
The \emph{overall intrinsic noise} is defined as
\begin{equation}
\N_i(E):=\sup_{\psi\in\hi_1}\N_i(E;\psi).
\end{equation}

The next definition applies to observables $E$ on $\R$ whose support is $\R$.\footnote{This 
means that for every interval $J$ there is a vector state $\psi$ such that $\prob_\psi^E(J)\ne 0$.}
The \emph{resolution width}  of $E$ (at confidence level
$1-\veps$) is  \cite{CaHeTo07}
\begin{equation}
\gamma_\veps(E):=\inf\{ d>0\,|\, \text{for all\ } x\in\R\,\text{there exists\ } 
\psi\in \hi_1\ \text{with} \ \prob_\psi^E(J_{x;d})\ge 1-\veps  \}.
\end{equation}
Here $J_{x;d}$ denotes the interval $[x-\frac d 2,x+\frac d 2]$.

We note that positive resolution width is a certain indicator that the observable $E$ is unsharp.
This measure describes the possibilities of concentrating the probability distributions to a fixed 
confidence level across all intervals. However, the requirement of vanishing resolution width 
does not single out sharp observables \cite{CaHeTo07}.

\subsection{Joint measurements}

Two observables $E_1$ and
$E_2$ on $\R$ are called \emph{jointly measurable} if there is an
observable $M$ on $\R^2$ such that
\begin{equation}
E_1(X)=M(X\times\R),\quad E_2(Y)=M(\R\times Y)
\end{equation}
for all (Borel) sets $X,Y$. Then $E_1$ and $E_2$ are the marginal observables $M_1$ and $M_2$ 
of the joint observable $M$. If either $E_1$ or $E_2$ is a sharp observable, then they are
jointly measurable exactly when they commute mutually. In that case, the unique
joint observable $M$ is determined by $M(X\times Y)=E_1(X)E_2(Y)$.
In general, the mutual commutativity of $E_1$ and $E_2$ is not a necessary
(although still a sufficient) condition for their joint measurability.

The above notion of joint measurability is fully supported by the quantum theory of measurement, 
which ensures that $E_1$ and $E_2$ are jointly measurable exactly when there is a measurement scheme which 
measures both $E_1$ and $E_2$ \cite{QTM}. 

Considering that the (sharp) position and momentum observables $\sfq$ and $\sfp$ do not commute with each other, we recover immediately the well-known fact that these observables have no joint observable, that is, they are not jointly measurable. This is a precise formulation of the no-go statement (B).

In preparation of developing a positive complementation to (B), we give an outline of the notion that sharp position  and momentum may be jointly measurable in an approximate sense. While there is no observable on phase space whose marginals coincide with $\sfq$ and $\sfp$, one can explore the idea that there may be observables $M$ on $\R^2$ whose marginals $M_1$ and $M_2$ are \emph{approximations} (in some suitably defined sense) of $\sfq$ and $\sfp$, respectively. Such an $M$ will be called an \emph{approximate joint observable} for $\sfq$ and $\sfp$.
An appropriate quantification of the differences between $M_1$ and $\sfq$ and between $M_2$ 
and $\sfp$ may serve as a measure of the \emph{(in)accuracy} of the joint approximate measurement 
represented by $M$.

\subsection{Sequential measurements}\label{sequential}

In order to analyze measurements of two observables $E_1$ and $E_2$ performed in immediate succession, it is necessary to take into account the influence of the first measurement on the object system.  The tool to  describe the state changes due to a measurement
is provided by the notion of an \emph{instrument}; see the Appendix for an explanation.

Let $\stfm_1$ be the instrument associated with a measurement of $E_1$, that is, $\stfm_1$ 
determines the probability $\tr{\rho E_1(X)}=\tr{\stfm_1(X)(\rho)}$ for every state $\rho$ and set $X$. 
The number $\tr{\stfm_1(X)(\rho)E_2(Y)}$ is the sequential joint probability that the measurement 
of $E_1$, performed on the system in state $\rho$, gives a result in $X$ and a subsequent 
measurement of $E_2$ leads to a result in $Y$. 
Using the dual instrument $\stfm^*_1$ (cf. the Appendix),
this probability
can be written as $\tr{\rho \stfm_1^*(X)(E_2(Y))}$. 
The map 
$(X,Y)\mapsto \tr{\rho \stfm_1^*(X)(E_2(Y))}$ is a probability bimeasure and therefore extends uniquely to a joint
probability for each $\rho$, defining thus a unique joint observable $M$ on $\R^2$ via
\begin{equation}
M(X\times Y)=\stfm_1^*(X)(E_2(Y)).\label{sequentialjoint}
\end{equation}
Its marginal observables are 
\begin{equation}
M_1(X)=\stfm_1^*(X)(E_2(\R))=E_1(X),\label{firstmarginal}
\end{equation}
\begin{equation}
M_2(Y)=\stfm_1^*(\R)(E_2(Y))=:E'_2(Y).\label{secondmarginal}
\end{equation}

\noindent
Thus the first marginal is the first-measured observable $E_1$ and the second marginal is a distorted version $E'_2$ of $E_2$.

This general consideration shows that one must expect that a measurement of an observable $E_1$
will \emph{disturb} (the distribution of) another observable $E_2$. In fact, it is a fundamental theorem
of the quantum theory of measurement that there is no nontrivial measurement without \emph{some}
state changes. In other words, if a measurement leaves \emph{all} states unchanged, then its statistics
will be the same for all states; in this sense there is no information gain without some disturbance.

If the first-measured observable $E_1$ is sharp, the distorted effects $\stfm_1^*(\R)(E_2(Y))$ must commute with $E_1(X)$ for all $X,Y$, whatever the
second observable $E_2$ is. 
Thus, if we consider a sequential measurement of the sharp position and momentum observables 
$\sfq$ and $\sfp$ as an attempted joint measurement,  we see that such an attempt is bound to
fail. If (say) one first measures position $\sfq$, with an instrument $\stfm_{\sfq}$,
then all distorted momentum effects $\sfp'(Y):=\stfm_{\sfq}^*(\R)(\sfp(Y))$ are functions of the position operator 
$\qhat$.  
In this sense, a measurement of sharp position completely destroys any information about the momentum distribution in the input state. This result formalizes the no-go statement (C).

The formulation of a positive complement to (C) is based on the idea that one may be able to control
and limit the disturbance due to a measurement of $\sfq$, by measuring 
an observable $\sfq'$
which is an approximation (in some sense) to $\sfq$. One can then hope to achieve that the distorted momentum 
$\sfp'$ is an approximation (in some sense) to $\sfp$. 
We note that this amounts to defining a sequential joint observable $M$ with marginals 
$M_1=\sfq'$ and $M_2=\sfp'$. Any appropriate quantification
 of the difference between 
$M_1$ and $\sfq$ is a measure of the inaccuracy of the first (approximate) position 
measurement; similarly
any appropriate quantification of the difference between $M_2$ and $\sfp$ is a measure of the disturbance of 
the momentum due to the position measurement. 

In this way the problem of defining measures of the disturbance of (say) momentum
due to a measurement of position has been reduced to defining the inaccuracy of 
the second marginal of a sequential joint measurement of first position and then momentum.

\subsection{On measures of inaccuracy}\label{obs-acc-dist}

The above discussion shows that it is the noncommutativity of observables such as
position and momentum which forces one to allow inaccuracies 
if  one attempts to make an approximate joint measurement of these observables.
This shows clearly that the required inaccuracies are of quantum-mechanical origin,
which will also become manifest in 
the models of approximate position measurements and phase space measurements
presented below. With this observation as proviso, we believe that it is acceptable to use 
the classical terms of measurement inaccuracy and error, particularly because their operational
definitions are essentially the same as in a classical measurement context.

In fact, every measurement, whether classical or quantum, is subject to \emph{noise}, which results
in a deviation of the actually measured observable $E_1$ from that
intended to be measured, $E$.
We will refer to this deviation and any measure of it as \emph{error} or 
\emph{inaccuracy}. In general there can be systematic errors, or \emph{bias},
leading to a shift of the mean values, and random errors, resulting in a broadening
of the distributions. 
Any  measure of measurement noise should be \emph{operationally
significant} in the sense that it should be determined
by the probability distributions  $\prob^E_\psi$ 
and  $\prob^{E_1}_\psi$. 

In the following we will discuss three different  approaches to quantifying measurement inaccuracy.

\subsubsection{Standard measures of error and disturbance}

Classical statistical analysis suggests the use of moments of probability distributions for
the quantification of error and disturbance in measurements. Thus, the standard approach
found in the literature of defining a measure of error is in terms of
the average deviation of the value of a readout observable of the measuring apparatus
from the value of the observable to be  measured approximately. 
If these observables are represented as selfadjoint operators $Z$ and $A$ (acting on the apparatus and the object Hilbert spaces), 
respectively, this \emph{standard error} measure is given as
the root mean square
\begin{equation}\label{classicalerror} 
\epsilon(Z,A,\psi):=
\ip{U(\psi\otimes\Psi)}{(Z-A)^2U(\psi\otimes\Psi)}^{1/2},
\end{equation}
where $U$ is the unitary map modelling the measuring interaction and $\Psi$ is the initial state of the apparatus. This  measure of error has been studied 
in recent years in the foundational context, for example, by Appleby \cite{Appleby98a,Appleby98b}, Hall \cite{Hall04} and Ozawa \cite{Ozawa04b}. 

If we denote by $E$ the observable actually measured by the given scheme, we define
the \emph{relative noise} operator,
\begin{equation}
N_r(E,A):=E[1]-A;
\end{equation}
the standard error can then be rewritten as \cite{Ozawa04b}
\begin{equation}\label{class-err}\begin{split}
\epsilon(E,A;\psi)^2&=\ip{\psi}{(E[1]-A)^2\psi}+\ip{\psi}{(E[2]-E[1]^2)\psi}\\
&=
\ip{\psi}{ N_r(E,A)^2\psi} + \ip{\psi}{ N_i(E)\psi}
\end{split}\end{equation}
(for any vector state $\psi$ for which the expressions are well-defined).
We note that $\eps(E,A;\psi)=0$ for all $\psi$ exactly when $E$ is sharp and $E[1]=A$.
The relative noise term cannot, in general, be determined from the statistics of measurements of $E$ and $A$ alone, so that the standard error measure $\eps(E,A;\psi)$ does not always satisfy the  
requirement of operational significance \cite{BuHeLa04}. However, we will encounter important 
cases where this quantity does turn out to be operationally well defined.

The standard error is a state-dependent quantity. This stands in contrast to the fact 
that estimates of errors obtained in a calibration process are meant to be applicable to a range of
states since in a typical measurement the state is unknown to begin with. In order to obtain state independent measures,
we define the \emph{global standard error} of an observable $E$ relative to $A$ as\footnote{This
definition was used by Appleby \cite{Appleby98b} for the special case of position and momentum observables.}
\begin{equation}
\eps(E,A):=\sup_{\psi\in\hi_1}\eps(E,A;\psi).
\end{equation}
We will say that $E$ is a \emph{standard approximation} to $A$ if $E$ has finite global
standard error relative to $A$.  This definition provides a possible criterion for selecting
joint or sequential measurements schemes as approximate joint measurements of 
$\sfq$ and $\sfp$; but it is not always possible to verify this criterion if the standard error
fails to be operationally significant.

\subsubsection{Geometric measure of approximation and disturbance}

Following the work of Werner  \cite{Werner04b},
we define a distance  $d(E_1,E_2)$ on the set of observables on $\R$.

We first recall that for any bounded measurable
function $h:\R\to\R$, the integral $\int_\R h\,dE$ defines (in the weak
sense) a bounded selfadjoint operator, which we denote by $E[h]$.
Thus, for any vector state $\psi$ the number $\ip{\psi}{E[h]\psi}=\int_\R h\,d\prob^E_\psi$ is well-defined.

Denoting by $\Lambda$ the set of bounded measurable functions $h:\R\to\R$ for which $|h(x)-h(y)|\leq |x-y|$, the distance between the observables $E_1$ and $E_2$ is  defined as
\begin{equation}\label{Werner-distance}
d(E_1,E_2) := \sup_{\psi\in\hi_1}\ \sup_{h\in\Lambda}\
\left|\ip{\psi}{\left( E_1[h]-E_2[h] \right) \psi}  \right|.
\end{equation}
This measure is operationally
significant, using only properties of the distributions to be compared. Furthermore, it is a global
measure in that it takes into account the largest possible deviations of the expectations
$\ip{\psi}{E_1[h]\psi}$ and $\ip{\psi}{E_2[h]\psi}$.
It gives a geometrically appealing quantification of
how well a given observable can be approximated by other observables.

We will say that an observable $E_1$ is a \emph{geometric approximation} to 
$E_2$ if $d(E_1,E_2)<\infty$.
We shall apply this condition of finite distance as a criterion for a joint or sequential measurement 
scheme to define an approximate joint measurement of $\sfq$ and $\sfp$. It is not clear
whether this criterion is practical since the distance is not related in any obvious way to concepts of measurement inaccuracy commonly applied in an experimental context.

\subsubsection{Error bars}

We now present a definition of measurement inaccuracy in terms of likely error intervals
that follows most closely the usual practice of calibrating measuring instruments. In the process of calibration of a measurement scheme, one seeks to obtain estimates of the
likely error and perhaps also the degree of disturbance that  the scheme contains. To estimate the error, one tests the device by applying it to a sufficiently large family of input 
states in which the observable one wishes to measure with this setup has fairly sharp values. The error is then characterized as an overall measure of the bias and 
the width of the output distribution across a range of input values. Error bars give the minimal average interval lengths that one has to allow to contain all output values with a given confidence level.

For each $\varepsilon\in(0,1)$, 
we say that an observable $E_1$ is an $\veps$-\emph{approximation} to a sharp observable $E$
if for all $\delta>0$ there is a positive number $w<\infty$ such that for  all $x\in\R$, $\psi\in\hi_1$,
the condition $\prob^E_\psi(J_{x;\delta})=1$ implies that $\prob_\psi^{E_1}(J_{x,w})\ge 1-\veps$. 
The infimum of all such $w$
will be called the \emph{inaccuracy} of $E_1$ with respect to
$E$ and will be denoted $\dele{E_1}$. Thus,
\begin{equation}\begin{split}
\dele{E_1}:=\inf\{w\,|\ &\text{for all\ }x\in\R,\,\psi\in\hi_1,\\
& \text{if\ } \prob^E_\psi(J_{x;\delta})=1
\ \text{then\ }\prob_\psi^{E_1}(J_{x,w})\ge 1-\veps
\}.
\end{split}\end{equation}
The inaccuracy describes the range within which the input values can be inferred 
from the output distributions, with confidence level $1-\varepsilon$, given initial localizations within 
$\delta$.
We note that the inaccuracy is an increasing function of $\delta$, so that we can
define the \emph{error bar width}\footnote{This definition and all subsequent results based on it
can be found in \cite{BuPe06}.} of $E_1$ relative to $E$:
\begin{equation}
\dee{E_1}:=\inf_\delta\dele{E_1}=\lim_{\delta\to 0}\dele{E_1}.
\end{equation}
If $\dee{E_1}$ is finite for all $\veps\in (0,\frac 12)$, we will say that $E_1$ approximates $E$ in the sense of \emph{finite error bars}.
We note that  the finiteness of 
either  $\eps(E_1,E)$ or  $d(E_1,E)$ implies the finiteness of $\dee{E_1}$.
Therefore, among the three measures of inaccuracy, the condition of finite error bars
gives the most general  criterion for selecting approximations of $\sfq$ and $\sfp$.

\section{From ``no joint measurements" to  approximate joint measurements}\label{mt-ur}

Position $\sfq$ and momentum $\sfp$ have no joint observable, they cannot be measured together.
However, one may ask for an approximate joint measurement, that is, for an observable
$M$ on $\R^2$ such that the marginals $M_1$ and $M_2$
are appropriate approximations of $\sfq$ and $\sfp$.
In this section we study two important cases and then consider the general situation.

\subsection{Commuting functions of position and momentum}\label{subsec:com}

The first approach is related with the fact that although $\sfq$ and
$\sfp$ are 
noncommutative, they do have commuting spectral
projections.
Indeed, let $\sfq^g$ be a function of $\sfq$, that is,
$\sfq^g(X)=\sfq(g^{-1}(X))$ for all (Borel) sets $X\subseteq\R$,  with $g:\R\to\R$
being a (Borel) function. Similarly, let $\sfp^h$ be a function of
momentum. The associated operators are $g(\qhat)$ and $h(\phat)$.
The following result, proved in \cite[Theorem
1]{BuScSc87} and in a more general setting in \cite{Ylinen89}, characterizes  the 
functions $g$ and $h$ for which
$\sfq^g(X)\sfp^h(Y) =\sfp^h(Y)\sfq^g(X)$ for all $X$ and $Y$.

\begin{theorem}\label{th:functions}
Let $g$ and $h$ be essentially bounded Borel functions such that
neither $g(\qhat)$ nor $h(\phat)$ is a constant operator. The
functions  $\sfq^g$ of position and  $\sfp^h$ of momentum
commute if and only if $g$ and $h$ are both periodic with
minimal positive periods $a,b$ satisfying $\frac{2\pi}{ab}\in \mathbb{N}$.
\end{theorem}

If $\sfq^g$ and $\sfp^h$ are commuting observables, then 
they have the joint observable $M$, with $M(X\times Y)= \sfq^g(X)\sfp^h(Y)$,
meaning that $\sfq^g$ and $\sfp^h$ can be measured jointly. The
price for this restricted form of joint measurability of position
and momentum as given by Theorem~\ref{th:functions} is that they
are to be coarse-grained by periodic functions $g$ and $h$ with
appropriately related minimal periods $a,b$.

The functions $g$ and $h$ can be chosen as characteristic functions
of appropriate periodic sets. This allows one to model a situation
known in solid state physics, where an electron in a crystal can be
confined arbitrarily closely to the atoms while at the same time its
momentum is localized arbitrarily closely to the reciprocal lattice
points.

Simultaneous localization of position and momentum in periodic
sets thus constitutes a sharp joint measurement of functions of
these observables. However, bounded functions $\sfq^f$ of 
$\sfq$ provide only very bad approximations to $\sfq$ since
$\eps(Q^f,Q)$, $d(\sfq^f,\sfq)$ and  $\mathcal{W}_{\veps,\delta}(Q^f,Q)$ 
are all infinite.  One also loses the characteristic
covariance properties of position and momentum.

\subsection{Uncertainty relations for covariant approximations of position and momentum}
Next we will discuss approximate joint measurements of position and momentum
based on smearings of these observables by means of convolutions. 

Let $\mu,\nu$ be probability measures on $\R$. We define observables 
$\sfq_\mu,\sfp_\nu$ via 
\begin{equation}\label{GT-marginals_gen}
\sfq_\mu(X)=\int_\R\mu(X-q)\,\sfq(dq),\qquad
\sfp_\nu(Y)=\int_\R\nu(Y-p)\,\sfp(dp).
\end{equation}
These observables have the same characteristic covariance properties as $\sfq,\sfp$ and
they are approximations in the sense that they have finite error bar widths relative to
$\sfq,\sfp$. Hence we call them approximate position and momentum.

For given $\sfq_\mu,\sfp_\nu$ we ask under what conditions they are jointly measurable,
that is, there is an observable $M$ on $\R^2$ such that $M_1=\sfq_\mu$ and $M_2=\sfp_\nu$.
In order to answer this question, we need to introduce the notion of covariant phase space 
observables.

Covariance is defined with respect to a unitary (projective) representation of phase space
translations in terms of the Weyl operators, defined for any phase space point $(q,p)\in\R^2$
via 
$W(q,p)=e^{\frac i{2\hbar} qp}\,e^{-\frac i\hbar q\phat}\,
e^{\frac i\hbar p\qhat}$.
An observable $M$ on $\R^2$ is a \emph{covariant phase space observable} if
\begin{equation}
W(q,p)M(Z)W(q,p)^*=M(Z+(q,p))
\end{equation}
for all $Z$. It is known\footnote{This result is due to Holevo \cite{Holevo79} and Werner \cite{Werner84}.
Alternative proofs with different techniques were recently given in
\cite{CaDeTo03} and \cite{KiLaYl06a}.} 
that each such observable is of the form
$G^T$, where
\begin{equation}\label{GW}
G^T(Z)=\frac 1{2\pi\hbar}\int_Z W(q,p)TW(q,p)^*\,dqdp,
\end{equation}
and $T$ is a unique positive operator of trace one.
The marginal observables $G^T_1$ and $G^T_2$ are of the form (\ref{GT-marginals_gen})
where $\mu=\mu_T=\prob^\sfq_{\varPi T\varPi^*}$ and $\nu=\nu_T=\prob^\sfp_{\varPi T\varPi^*}$
and $\varPi$ is the parity operator, $\varPi\psi(x)=\psi(-x)$.

Our question is answered by the following fundamental theorem which was proven in the
present form  \cite[Proposition 7]{CaHeTo04} as a direct development of the
work of \cite{Werner04b}. 
\begin{theorem}\label{th:covariance}
An approximate position $\sfq_{\mu}$ and an approximate momentum
$\sfp_{\nu}$ are jointly measurable  if and only if they have a
covariant joint observable $G^T$. This is the case exactly when there is a positive 
operator $T$ of trace equal to 1 such that $\mu=\mu_T$, 
$\nu=\nu_T$.
\end{theorem}

We next quantify the necessary trade-off in the quality of the approximations of $\sfq,\sfp$ by
$G^T_1,G^T_2$, using the three measures of inaccuracy introduced above. First we state
uncertainty relations for the measures of intrinsic unsharpness. 
The probability distributions associated with these marginals of $G^T$ are the convolutions 
$\prob^{\sfq}_\psi*\mu_T$ and $\prob^{\sfp}_\psi*\nu_T$; this indicates that
statistically independent noise is added to the distributions of sharp position and momentum.
In accordance with this fact, the standard deviations are obtained via the sums of variances,
\begin{equation}\label{marg-var}
\Delta(G^T_1,\psi)^2=\Delta(\sfq,\psi)^2+\Delta(\mu_T)^2,\qquad
\Delta(G^T_2,\psi)^2=\Delta(\sfp,\psi)^2+\Delta(\nu_T)^2.
\end{equation}
The noise interpretation is confirmed by a determination of the intrinsic noise operators of $G^T_1,G^T_2$, which are well-defined whenever the operator $T$ 
is such that $\hat Q^2\sqrt{T}$ and $\hat P^2\sqrt{T}$ are Hilbert-Schmidt operators
\cite[Theorem 4]{KiLaYl06b}; in that case one obtains
\begin{equation}
N_i(G^T_1)= \Delta(\mu_T)^2\,I,
\quad
N_i(G^T_2) =\Delta(\nu_T)^2\,I. 
\end{equation}
Since $T$, and with it $\varPi T\varPi^*$, has the properties of a state, the uncertainty relation (\ref{p-ur}) applies to the probability measures $\mu_T$ and $\nu_T$, giving the following uncertainty relation for intrinsic noise, valid for any $G^T$:
\begin{equation}\label{noise-ur}
\N_i(G^T_1)\cdot\N_i(G^T_2)=\Delta(\mu_T)^2\cdot\Delta(\nu_T)^2\geq \frac{\hbar^2}{4}.
\end{equation}

Equations (\ref{marg-var}) and (\ref{noise-ur}) yield the following version of state-preparation uncertainty relation with respect to  $G^T$, which also reflects the presence of the intrinsic noise:
\begin{equation}\label{phasespaceur}
\Delta(G^T_1,\psi)\cdot\Delta(G^T_2,\psi)\geq\hbar.
\end{equation}

The resolution widths of $G^T_1,G^T_2$ are given by
\begin{equation}
\gamma_{\veps_1}(G^T_1)=W_{\veps_1}(\mu_T),\quad
\gamma_{\veps_2}(G^T_2)=W_{\veps_2}(\nu_T),
\end{equation}
so that the uncertainty relation for overall widths then entails:
\begin{equation}
\gamma_{\veps_1}(G^T_1)\cdot \gamma_{\veps_2}(G^T_2)
=W_{\veps_1}(\mu_T)\cdot W_{\veps_2}(\nu_T)\ge 2\pi\hbar\cdot (1-\veps_1-\veps_2)^2.
\end{equation}

The standard errors are
\begin{equation}
\eps(G^T_1,\sfq;\psi)^2=\big(\mu_T[1]\big)^2+\Delta(\mu_T)^2,\qquad
\eps(G^T_2,\sfp;\psi)^2=\big(\nu_T[1]\big)^2+\Delta(\nu_T)^2,
\end{equation}
and the inequality
\begin{equation}\label{h2again}
\eps(G^T_1,\sfq)\cdot\eps(G^T_2,\sfp)\geq\frac{\hbar}{2}
\end{equation}
holds as an immediate consequence of the noise uncertainty relation (\ref{noise-ur}). 

The distances of $G^T_1,G^T_2$ from $\sfq,\sfp$ are 
\begin{equation}
d(G^T_1,\sfq)=\int|q|\mu_T(dq),\qquad d(G^T_2,\sfp)=\int|p|\nu_T(dp),
\end{equation}
and they satisfy the trade-off inequality
\begin{equation}\label{Werner_ur}
d(G^T_1,\sfq)\,\cdot d(G^T_2,\sfp)\geq C\hbar,
\end{equation}
where the value of the constant $C$ can be numerically determined
as  $C\approx 0.3047$ \cite{Werner04b}. 
There is a unique covariant joint observable $G^T$ attaining the lower bound in 
(\ref{Werner_ur}), but the optimizing operator $T=\kb{\eta}{\eta}$ is not given by the 
oscillator ground state \cite[Section 3.2]{Werner04b}.

Finally considering the error bar widths of $G^T_1,G^T_2$ relative to $\sfq,\sfp$, one finds:
\begin{equation}
\deq{G^T_1}\ge W_{\veps_1}(\sfq,T),\qquad \dep{G^T_2}\ge W_{\veps_2}(\sfp,T). 
\end{equation}
Therefore, (\ref{ow-ineq}) implies that 
\begin{equation}\label{barineq}
\deq{G^T_1}\cdot\dep{G^T_2}\geq {2\pi\hbar}\cdot\left( 1-\veps_1-\veps_2 \right)^2.
\end{equation}
We note that  error bar widths in this inequality are always finite, in contrast to 
the standard errors or distances, which are infinite for some $G^T$.

The existence of covariant phase space observables $G^T$ establishes the positive complement
to the no-go statement (B). We have given Heisenberg uncertainty relations for the \emph{necessary}
inaccuracies in the approximations of $\sfq,\sfp$ by means of the marginal observables
$G^T_1$, $G^T_2$. For each pair of values of the inaccuracies allowed by these uncertainty relations
there exists a $G^T$ which realizes these values. This confirms the
\emph{sufficiency} of the inaccuracy relations for the existence of an approximate joint measurement of 
position and momentum, in the form of a covariant joint observable.

There is a (perhaps unexpected) reward for the positive attitude that led to the search 
for approximate joint measurements of position and momentum: the family of 
covariant phase space observables $G^T$ contains the important class of  
\emph{informationally complete} phase space observables. An example is
given by the choice $T=\kb{\eta_{a,0}}{\eta_{a,0}}$.

\subsection{Uncertainty relations for general approximate joint measurements}

While the uncertainty relations are necessary for the inaccuracies inherent in
jointly measurable covariant approximations $\sfq_\mu$ and $\sfp_\nu$, there remains
the possibility that one can overcome the Heisenberg limit by some clever choice of
non-covariant approximations of $\sfq$ and $\sfp$. 
Here we show that this possibility is ruled out. It follows that
covariant phase space observables constitute the optimal class of approximate
joint observables for position and momentum.

Let $M$ be an observable on $\R^2$. It was shown by Werner \cite{Werner04b} that 
if $M_1,M_2$ have finite distances from $\sfq,\sfp$, respectively, then
there is a covariant phase space observable $G^T$ associated with $M$ with the following property:
 $d(M_1,\sfq)\ge d(G^T_1,\sfq)$ and
$d(M_2,\sfp)\ge d(G^T_2,\sfp)$. The same kind of argument can be
carried out in the case of the global standard error and the error bar width\footnote{See \cite{BuPe06}; inequality (\ref{h2again_gen}) was deduced by different methods in \cite{Appleby98b}.} so that 
the inequalities (\ref{h2again}), (\ref{Werner_ur}) and (\ref{barineq}) 
entail the universally valid Heisenberg uncertainty relations
\begin{equation}
\eps(M_1,\sfq)\cdot\eps(M_2,\sfp)\geq
\frac{\hbar}{2}, \label{h2again_gen}
\end{equation}
\begin{equation}
d(M_1,\sfq)\,\cdot d(M_2,\sfp)\geq 
C\hbar, \label{Werner_ur_gen}
\end{equation}
\begin{equation}
\deq{M_1}\cdot\dep{M_2}\geq  
{2\pi\hbar}\cdot\left( 1-\veps_1-\veps_2 \right)^2.\label{barineq_gen}
\end{equation}
We propose the conjecture that these inaccuracy relations can be complemented with equally general trade-off relations for the intrinsic noise and resolution width of the marginals of 
an approximate joint observable of $\sfq,\sfp$:
\begin{equation}
\N_i(M_1)\cdot\N_i(M_2)\geq \frac{\hbar}{2},
\end{equation}
\begin{equation}
\gamma_{\veps_1}(M_1)\cdot \gamma_{\veps_2}(M_2)
\ge 2\pi\hbar\cdot (1-\veps_1-\veps_2)^2.
\end{equation}

Considering now examples  of noncovariant observables on phase space, 
we recall first that the commutative 
observable $M$ on $\R^2$ of subsection \ref{subsec:com} has marginals with infinite error bars.
Here we give an example of an observable  $M$ on phase space which is not covariant but is still
an approximate joint observable for $\sfq,\sfp$. Let $G^T$ be a
covariant phase space observable and define $M:=G^T\circ\gamma^{-1}$, where $\gamma(q,p):=
(\gamma_1(q),\gamma_2(p))$. We assume that $\gamma_1,\gamma_2$ are strictly increasing 
continuous functions such that $\gamma_1(q)-q$ and $\gamma_2(p)-p$ are bounded functions.
Then it follows  that the marginals $M_1^\gamma=G^T_1\circ\gamma_1^{-1}$ and
$M_2^\gamma=G^T_2\circ\gamma_2^{-1}$ have finite error bars with respect to $\sfq,\sfp$. If 
$\gamma$ is a nonlinear function then $M$ will not be covariant.

\section{From ``no measurement without disturbance" to sequential joint measurements}\label{disturb}

As concluded in Subsection~\ref{sequential} there is no way to determine the (sharp) position and momentum observables in a sequential measurement. We show now that  there are sequential 
measurements which are approximate simultaneous determinations of position and momentum.
As discussed above, the inaccuracy of the second measurement defines an operational measure of the 
disturbance of momentum due to the first, approximate measurement of position. It therefore follows that 
for any sequential joint observable $M$ on $\R^2$ the inaccuracies satisfy the trade-off relations 
(\ref{h2again_gen}), (\ref{Werner_ur_gen}) and (\ref{barineq_gen}) and, moreover, these relations
constitute now the long-sought-for inaccuracy-disturbance trade-off relations. 

Insofar as there are sequential measurement schemes in which these error and disturbance
measures are finite, we have thus established the positive complement to the no-go statement (C):
the associated sequential joint observable constitutes an approximate joint measurement, so
that it is indeed possible to limit the disturbance of the momentum by allowing the position 
measurement to be only approximate.

The existence of sequential measurements of approximate position and momentum can be
demonstrated by means of the ``standard model" of an unsharp position measurement
introduced by von Neumann \cite{MGQ}. In this model, the position of the
object is measured by coupling it to the momentum $\sfp_p$ of the probe system
via $U=e^{-(i/\hbar)\lambda \qhat\otimes \phat_p}$, and using the position $\sfq_p$ of
the probe as the readout observable. If $\Psi_p$ is the initial probe state, then the instrument of the measurement can be written in the form\footnote{In formula (\ref{standardQ}) we assume that the probe state $\Psi_p$ is a bounded function. As shown in \cite[Section 6.3]{CaHeTo07}, this assumption can be lifted by defining the instrument in a slightly different way.}

\begin{equation}\label{standardQ}
\stfm(X)(\rho)=\int_X K_q\rho K_q^*\,dq,
\end{equation}
with $K_q$ denoting the multiplicative operator $(K_q\psi)(x)=
\sqrt{\lambda}\ \Psi_p(\lambda(q-x))\psi(x)$. The approximate
position realized by this measurement is $\sfq_\mu$,
where $\mu$ is now the probability measure with distribution function 
$\lambda|\Psi_p(-\lambda x)|^2$.

Suppose now that one is carrying out first an approximate position measurement, 
with the instrument (\ref{standardQ}), and then a sharp momentum measurement.
As shown by Davies \cite{Davies70}, this defines a unique sequential joint observable $M$,
in fact, a covariant phase space observable with marginals 
\begin{equation}\label{seq-marginals}
M_1(X)=\stfm^{*}(X)(\sfp(\R))=\sfq_\mu(X),\qquad 
M_2(Y)=\stfm^{*}(\R)(\sfp(Y))= \sfp_\nu(Y).
\end{equation}
Here the distorted momentum is 
$\sfp_\nu$, where $\nu$ is the probability measure with the distribution $\frac
1\lambda|\widehat\Psi_p(-\frac p\lambda)|^2$.
It is obvious that $\mu=\mu_T$ and $\nu=\nu_T$, where 
$T=\kb{\Psi^{(\lambda)}}{\Psi^{(\lambda)}}$ with 
$\Psi^{(\lambda)}(q)=\sqrt{\lambda}\Psi_p(\lambda q)$. This makes it manifest that $M$
obeys the uncertainty relations 
(\ref{h2again_gen}), (\ref{Werner_ur_gen}) and (\ref{barineq_gen}), here in their
double role as accuracy-accuracy and accuracy-disturbance trade-off relations.

\section{Illustration: the Arthurs-Kelly model}

The best studied model of a joint measurement of position and
momentum is that of Arthurs and Kelly \cite{ArKe65}. In this model,
a quantum object is coupled with two probe systems which are then
independently measured to obtain information about the object's
position and momentum respectively. Arthurs and Kelly showed that
this constitutes a simultaneous measurement of position and momentum
in the sense that the distributions of the outputs reproduce the
quantum expectation values of the object's position and momentum.
They also derived the uncertainty relation for the spreads of the
output statistics corresponding to our Eq.~(\ref{phasespaceur}). As
shown in \cite{Busch82}, the model also satisfies the more stringent
condition of an approximate joint measurement, that the output statistics
determine a covariant phase space observable whose marginals are
smeared versions of position and momentum. This work also extended the model to a large
class of probe input states (Arthurs and Kelly only considered
Gaussian probe inputs), which made it possible to analyze the origin
of the uncertainty relation for the measurement accuracies and
identify the different relevant contributions to it. This will be
described briefly below. For a detailed derivation of the induced
observable and the state changes due to this measurement scheme, see
\cite{Busch82} and \cite[Chapter 6]{OQP}. Further illuminating
investigations of the Arthurs-Kelly model can be found, for
instance, in \cite{Stenholm92} and \cite{Raymer94}.

The Arthurs-Kelly model is based on the von Neumann model of an approximate
position measurement introduced in Sec.~\ref{disturb}. The position $\qhat$
and momentum $\phat$ of the object are coupled with the position $\qhat_1$ and
momentum $\phat_2$ of two probe systems, respectively, which serve as the
readout observables. Neglecting the free evolutions of the three systems the
combined time evolution is described by the measurement coupling
\begin{equation}
U=\exp\left(-\frac {i\lambda}\hbar \qhat\otimes\phat_1\otimes I_2+\frac
{i\kappa}\hbar\phat\otimes I_1\otimes\qhat_2\right).
\end{equation}
If $\psi$ is
an arbitrary input (vector) state of the object, and $\Psi_1,\Psi_2$ are the
fixed initial states of the probes (given by suitable smooth functions, with
zero expectations for the probes' positions and momenta), the probabilities for
values of $\qhat_1$ and $\phat_2$ to lie in the intervals $\lambda X$ and $\kappa
Y$, respectively, determine a covariant phase space observable $G^T$ of the
form (\ref{GW}) via
\begin{equation}
\langle\psi|G^T(X\times Y)|\psi\rangle:=
\langle U\psi\otimes\Psi_1\otimes\Psi_2|I\otimes\sfq_1(\lambda X)\otimes\sfp_2(\kappa Y)
|U\psi\otimes\Psi_1\otimes\Psi_2\rangle.
\end{equation}
The variances of the accuracy measures $\mu,\nu$ associated with the marginals $\sfq_\mu,\sfp_\nu$
of $G^T$ can readily be computed:
\begin{equation}\begin{split}
\Delta(\mu)^2&=\frac 1{\lambda^2}\Delta(\qhat_1,\Psi_1)^2+\frac{\kappa^2}4\Delta(\qhat_2,\Psi_2)^2,\\
\Delta(\nu)^2&=\frac 1{\kappa^2}\Delta(\phat_2,\Psi_2)^2+\frac{\lambda^2}4\Delta(\phat_1,\Psi_1 )^2.
\end{split}\end{equation}
If the two measurements did not disturb each other, only the first terms on the right hand
sides would appear; the second terms are manifestations of the presence of the other
probe and its coupling to the object. Since the
observable defined in this measurement scheme is a covariant phase space observable,
 it follows immediately that the accuracy measures satisfy the trade-off relation
(\ref{noise-ur}), $\Delta(\mu)\Delta(\nu)\ge\hbar/2$. It is nevertheless instructive to verify
this explicitly by evaluating the product of the above expressions:
\begin{equation}\begin{split}
\Delta(\mu)^2&\Delta(\nu)^2=\mathcal{Q}+\mathcal{D},\\
\mathcal{Q}:=&\frac 14 \Delta(\qhat_1,\Psi_1)^2\Delta(\phat_1,\Psi_1)^2+
\frac 14 \Delta(\qhat_2,\Psi_2)^2\Delta(\phat_2,\Psi_2)^2\ge\frac{\hbar^2}8\\
\mathcal{D}:=&\frac 1{(\lambda\kappa)^2}\Delta(\qhat_1,\Psi_1)^2\Delta(\phat_2,\Psi_2)^2+
\frac{(\lambda\kappa)^2}{16}\Delta(\qhat_2,\Psi_2)^2\Delta(\phat_1,\Psi_1)^2\\
&\hspace{1.5cm}\ge
\frac{\hbar^2}{16}\left(x+\frac 1x\right)\ge\frac{\hbar^2}8,\\
&\text{where\ }x:=\frac{16}{(\lambda\kappa\hbar)^2}\Delta(\qhat_1,\Psi_1)^2\Delta(\phat_2,\Psi_2)^2.
\end{split}\end{equation}
Here we have repeatedly used the uncertainty relations for the probe systems,
$\Delta(\qhat_k,\Psi_k)\Delta(\phat_k,\Psi_k)\ge\hbar/2$.

It is evident that there are two independent sources of inaccuracy in this
joint measurement model. Indeed, each of the terms $\mathcal{Q}$ and
$\mathcal{D}$ alone would suffice to guarantee an absolute positive lower bound
for the inaccuracy product. The first term, $\mathcal{Q}$, is composed of two
independent terms which reflect the \emph{quantum nature} of the probe systems;
there is no trace of a mutual influence of the two measurements being carried
out simultaneously. This feature is in accordance with Bohr's argument
concerning the possibilities of measurement, which he considered limited due to
the quantum nature of parts of the measuring setup (the probe systems).

By contrast, the term $\mathcal{D}$ reflects the mutual \emph{disturbance} of
the two measurements as it contains the coupling parameters and product
combinations of variances associated with both probe systems. This feature of
the mutual disturbance of measurements was frequently highlighted by Heisenberg
in thought experiments aiming at joint or sequential determinations of the
values of position and momentum.

A suitable modification of the measurement coupling $U$ leads to a model that
can be interpreted as a sequential determination of position and momentum.
Consider the unitary operator, dependent on the additional real parameter
$\gamma$,
\begin{equation}
U^{(\gamma)}=\exp\left(-\frac {i\lambda}\hbar\qhat\otimes\phat_1\otimes I_2+
\frac {i\kappa}\hbar\phat\otimes I_1\otimes \qhat_2-\frac
{i\gamma\lambda\kappa}{2\hbar} I\otimes\phat_1\otimes\qhat_2\right).
\end{equation}
The Baker-Campbell-Hausdorff decomposition of this coupling yields
\begin{equation}\begin{split}
U^{(\gamma)}=&\exp\left(-(\gamma+1)\frac i{2\hbar}\lambda\kappa I\otimes\phat_1\otimes\qhat_2\right)
\times\\
&\hspace{1cm}\times\exp\left(-\frac i\hbar\lambda\qhat\otimes\phat_1\otimes I_2\right)
\exp\left(\frac i\hbar\kappa\phat\otimes I_1\otimes \qhat_2\right).
\end{split}\end{equation}
It turns out that this coupling defines again a covariant phase space observable.
The variances of the inaccuracy measures $\mu_\gamma$, $\nu_\gamma$ associated
with the marginals are given as follows:
\begin{equation}\begin{split}
\Delta(\mu_\gamma)^2&=\frac 1{\lambda^2}\Delta(\qhat_1,\Psi_1)^2+
(\gamma-1)^2\frac{\kappa^2}4\Delta(\qhat_2,\Psi_2)^2,\\
\Delta(\nu_\gamma)^2&=\frac 1{\kappa^2}\Delta(\phat_2,\Psi_2)^2+
(\gamma+1)^2\frac{\lambda^2}4\Delta(\phat_1,\Psi_1)^2.
\end{split}\end{equation}
These accuracies still satisfy the uncertainty relation (\ref{noise-ur}), but
this time the contributions corresponding to $\mathcal{Q}$ and $\mathcal{D}$
will both depend on the coupling parameters unless $\kappa=0$. In particular,
it does not help to make the coupling look like that of a sequential
measurement, by putting $\gamma=-1$. In that case, $\Delta(\nu_{-1})$ is the
accuracy of an undisturbed momentum measurement, and $\Delta(\mu_{-1})$
contains a term which reflects the disturbance of the subsequent position
measurement through the momentum measurement. The disturbance of the position
measurement accuracy is now given by $\kappa\Delta(\qhat_2,\Psi_2)$, and together
with the momentum inaccuracy it satisfies the uncertainty relation
\begin{equation}
\Big[\frac 1{\kappa^2}\Delta(\phat_2,\Psi_2)^2\Big]\,
\left[\kappa^2\Delta(\qhat_2,\Psi_2)^2\right]\ge\frac{\hbar^2}4.
\end{equation}

\section{On experimental implementations and tests of the uncertainty principle}\label{expt}

\begin{quote}
``Turning now to the question of the empirical support [for the
uncertainty principle], we unhesitatingly declare that rarely in the
history of physics has there been a principle of such universal
importance with so few credentials of experimental tests."
\cite[p.~81]{PQM}
\end{quote}

This assessment was written by the distinguished historian of physics Max
Jammer at a time when studies  of phase space observables based on
positive operator measures were just beginning. He qualifies it with
a survey of early proposed and actual tests of the preparation
uncertainty relation, and he refers to some early model studies of
joint measurements, the first of which being that by Arthurs and Kelly \cite{ArKe65}.

Jammer's verdict still holds true today. There are surprisingly few
publications that address the question of experimental tests  of the
uncertainty principle. Some of these report confirmations of the
uncertainty principle, while a few others predict or suggest
violations. We will briefly comment on some of this work below.

\subsection{Tests of preparation uncertainty relations}

The most commonly cited version of uncertainty relation is the  preparation
relation, usually in the familiar version in terms of standard deviations. Confirmations
of this uncertainty relation have been reported by Shull \cite{Shull69} for a
single-slit diffraction experiment with neutrons, by Kaiser \emph{et al} \cite{KaWeGe83} and 
Klein \emph{et al} \cite{KlOpHa83} in
neutron interferometric experiments, and more recently by Nairtz \emph{et al} \cite{NaArZe02} in a
slit experiment for fullerene molecules.

In these slit diffraction and interferometric experiments, typical
measures used for the width of the spatial wave function are the
slit width and slit separation, respectively. The width of the
associated momentum wave function is given in terms of the width at
half height of the central peak. It must be noted that in the
mathematical modeling  of single slit diffraction, the standard
deviation of the momentum distribution is infinite. Hence it is
indeed necessary to use another, operationally significant measure of the width of
that distribution. There does not seem to be a universally valid
uncertainty relation involving width at half height (in short, half
width), but the authors of these experiments make use of a Gaussian
shape approximation of the central peak, which is in agreement with
the data within the experimental accuracy. This allows them to
relate the half widths to standard deviations and confirm the
correct lower bound for the uncertainty product.

A model independent and thus more direct confirmation of the uncertainty
principle can be obtained if the widths of the position and momentum
distributions are measured in terms of the overall width defined in
Eq.~(\ref{overall-width}). It is likely that the data collected in these
experiments contain enough information to determine these overall widths for
different levels of total probability $1-\varepsilon_1$ and $1-\varepsilon_2$.
In the case of the neutron interference experiment, it was pointed out by
Uffink \cite{Uffink85} that a more stringent relation is indeed at stake, namely, a
trade-off relation, introduced by Uffink and Hilgevoord \cite{UfHi85}, between the overall width of
the position distribution and the fine structure width (mean peak width) of the
momentum distribution.

It should be noted that these experiments do not, strictly speaking, constitute \emph{direct}
tests of the uncertainty relations for position and momentum observables. While the position
uncertainty, or the width of the position distribution, is determined as the width of the slit,
the momentum distribution is inferred from the measured position distribution at
a later time, namely when the particles hit the detection screen. This inference is based on the
approximate far-field description of the wave function (Fraunhofer diffraction in optics), and
is in accordance with the classical, geometric interpretation of momentum as mass times velocity.
Thus, what is being tested is the uncertainty relation along with the free Schr\"odinger evolution
and the Fourier-Plancherel connection between position and momentum.

An alternative interpretation can be given in the Heisenberg
picture, noting that the operators $\qhat$, $\phat':=m\qhat(t)/t$
are canonically conjugate, given the free evolution $\qhat(t)=\qhat+
\phat t/m$. Here $m$ is the mass of the particle, and $t$ is the
time of passage of the wave packet from  the slit to the detection
screen. (If the distance between the slit and the detection screen
is large compared to the longitudinal width of the wave packet, the
time $t$ is fairly well defined.) The width of the distribution of
$\qhat$ is determined by the preparation (passage through the slit),
and the distribution of $\phat'$ is measured directly.

\subsection{On implementations of joint and sequential measurements}

To the best of our knowledge, and despite some claims to the
contrary, there is presently no experimental realization of a joint
measurement of position and momentum. Thus there can as yet be no
question of an experimental test of the uncertainty relation for
inaccuracies in joint measurements of these quantities. But there
are reports on the successful experimental implementation of joint
measurements of canonically conjugate quadrature components of
quantum optical fields using multiport homodyne detection.

There seem to be several communities in quantum optics and optical
communication where these implementations were achieved
independently. The experiment of Walker and Carrol \cite{WaCa86} is perhaps the first
realization, with a theoretical analysis  by Walker \cite{Walker87} yielding
the associated phase space observable. This seems to  have been
anticipated theoretically by Yuen and Shapiro \cite{YuSh80}.  See also 
Lai and Haus \cite{LaHa89}
for a review. A more recent claim of a quantum optical realization
of a joint measurement  was made by Beck \emph{et al} \cite{BeDoWa01}. It must be
noted that in these works it is not easily established (in some
cases for lack of sufficiently detailed information) whether the
implementation criterion is merely that of reproducing the first
moments of the two quadrature component statistics, or whether in
fact the statistics of a joint observable have been measured.

By contrast, Freyberger \emph{et al} \cite{FrVoSc93}, \cite{FrSc93} and 
Leonhardt \emph{et al} \cite{LePa93}
\cite{LeBoPa95} showed that the eight port homodyne schemes for
phase difference measurements carried out by Noh \emph{et al} \cite{NoFoMa91},
\cite{NoFoMa92} yield statistics that approach the Q-function of the
input state for a suitable macroscopic coherent state preparation of
the local field mode. This is manifestly a realization of a joint
observable. A simple analysis is given in \cite[Sec. VII.3.7.]{OQP}.

Turning to the question of position and momentum proper, the
Arthurs-Kelly model is particularly well suited to elucidate the
various aspects of the uncertainty principle for joint and, as we
have seen, sequential joint measurements of approximate position and
momentum. However, it is not clear whether and how an experimental
realization of this scheme can be obtained. Apart from the  quantum
optical realizations of joint measurements of conjugate quadrature
components, there are a few proposals of realistic schemes for
position and momentum, e.g., \cite{Royer85}, \cite{ToStJe95}, and
\cite{PoTaWi97} mainly in the context of atom optics. In the latter
two models the probe systems are electromagnetic field modes, and
the readout probe observables are suitable phase-sensitive
quantities. The measurement coupling differs from the Arthurs-Kelly
coupling in accordance with the different choice of readout
observables.

The experimental situation regarding the
inaccuracy-versus-disturbance relation is far less well developed.
This is probably because, as we have seen above, rigorous,
operationally relevant formulations of such a relation had not been
found until recently. Apart from some model considerations of the
kind considered here in Sec.~III there seems to be no experimental
investigations of accuracy-disturbance trade-off relations.

\subsection{On some alleged violations of the uncertainty principle}

Throughout the history of quantum mechanics, the joint measurement
uncertainty relation has been the subject of repeated challenges. There
are two lines of argument against it which start from logically contrary
premises. The conclusion is, in either case, that only the preparation
uncertainty relation is tenable (as a statistical relation) within quantum mechanics.

The first argument against the joint measurement relation was based
on the claim that there is no provision for a notion of joint
measurement within quantum mechanics. Based on a careful assessment
of the attempts existing at the time, Ballentine \cite{Ballentine70}
concludes that a description of joint measurements of position and
momentum in terms of joint probabilities could not be obtained
without significant modifications or extensions of the existing
theory. Here we have shown that the required
modification was the introduction of positive operator measures and
specifically phase space observables, which is entirely within the
spirit of the traditional formulation of quantum mechanics; it
amounts merely to a completion of the set of observables.

The second argument was based on the claim that joint measurements
of position and momentum are in fact possible with arbitrary
accuracy, and its authors, among them Karl Popper and Henry Margenau, attempted
to demonstrate their claim by means of appropriate experimental
schemes.

Popper \cite{Popper34} conceived a joint
measurement scheme that was  based on measurements of entangled
particle pairs. That this proposal was flawed and untenable was
immediately noted by von Weizs\"acker \cite{Weizsacker34}. While
Popper later accepted this criticism, he suggested \cite[footnote on
p.~15]{QTSP} that his example may nevertheless have inspired
Einstein, Podolsky and Rosen  \cite{EiPoRo35} to conceive their famous thought experiment. 
 In fact, this experiment can be construed as a
scheme for making a joint measurement of the position and
momentum of a particle that is entangled with another particle in a
particular state: \emph{provided} that Einstein, Podolsky and
Rosen's assumption of local realism is tenable, a measurement of the
position of the latter particle allows one to infer the position of
the first particle without disturbing that particle in any way. At
the same time, one can then also measure the position of the first
particle.

It would follow that the individual particle has definite values of
position and  momentum while quantum mechanics provides only an
incomplete, statistical description. However, it a well-known
consequence of arguments such as the Kochen-Specker-Bell theorem
\cite{KoSp67,Bell66} and Bell's theorem \cite{Bell64} that such
value assignments are in contradiction with quantum mechanics.
Moreover, this contradiction has been experimentally confirmed in
the case of Bell  's inequalities, and these tests turned out in
favor of quantum mechanics.

Another proposal of a joint determination of arbitrarily sharp
values of the position and momentum of a quantum particle was made
by Park and Margenau \cite{PaMa68} who considered the time of flight
determination of velocity. As shown in a quantum mechanical analysis
in \cite{BuLa84}, this scheme is appropriately understood as a
sequential measurement of first sharp position and then sharp
momentum, and does therefore not constitute even an approximate joint 
measurement of position and momentum.
But Park and Margenau are only interested in demonstrating
that it is possible to ascribe arbitrarily sharp values of position and momentum
to a single system at the same time.

An analogous situation arises in the slit experiment, where one could
formally infer arbitrarily sharp values for the transversal momentum component
from the bundle of geometric paths from any location in the slit to the detection point.
This bundle is arbitrarily narrow if the separation between slit and detection screen is
made sufficiently large. Thus the width of the spot on the detection screen and the width
of the possible range of the inferred momentum value can be made small enough so
that their product is well below the order of $\hbar$.

In both situations, the geometric reconstruction of a momentum
value from the two position determinations at different times, which is
guided by classical reasoning, constitutes an inference for the time between the
two measurements and cannot be used to infer momentum
distributions in the state before the measurement or to predict the outcomes of
future measurements. Hence such values are purely formal and of no operational significance.
One could be inclined to follow Heisenberg who noted in his 1929 Chicago lectures \cite[p. 25]{PPQT}
that he regarded it as a matter of taste whether one considers such value assignments to past events as meaningful.

However, it has been shown, by an extension of the quantum mechanical language to incorporate
propositions about past events, that hypothetical value assignments to past events lead to
Kochen-Specker type contradictions. This result was obtained by Quadt \cite{Quadt88} in his diploma thesis written under P.~Mittelstaedt's supervision at the University of Cologne in 1988;
the argument is sketched in \cite{Quadt89}.

Popper returned to the subject many years later 
\cite[pp.~27-29]{QTSP} with a novel experimental proposal with which he aimed
at testing (and challenging) the Copenhagen interpretation. In a
subsequent experimental realization it is reported that the outcome seems to confirm Popper's prediction, thus amounting to an apparent violation of the preparation uncertainty relation.

In Popper's new experiment, EPR-correlated pairs of quantum particles
are emitted from a source in opposite directions, and then each
particle passes through a slit, a narrow one on one side, the one on
the opposite side of wide opening. The particles are then recorded
on a screen on each side. Popper predicts that independent
diffraction patterns should build up on each side, according to the
appropriate slit width; according to Popper, the Copenhagen
interpretation should predict that the particle passing through the
wider slit actually shows the same diffraction pattern as the other
particle. In the extreme of no slit on one side, this would still be
the case. Popper's interpretation of his experiment as a test of the
Copenhagen interpretation was criticized soon afterwards, see, e.g.,
the exchange in \cite{CoLo87a,Popper87,CoLo87b} or \cite{Sudbery85}.

The experimental realization of Popper's experiment by Kim and Shih
\cite{KiSh99} shows, perhaps at first surprisingly, a behavior in line with
Popper's prediction. Moreover, taking the width of the ``ghost
image" of the first, narrow slit at the side of the second particle
(confirmed in \cite{PiShStSe95}) as a measure of the position uncertainty
of the second particle, then this value together with the inferred width of the
momentum distribution form a product smaller than allowed by the
preparation uncertainty relation. Kim and Shih hasten to assert that
this result does not constitute a \emph{violation} of the
uncertainty principle but is in agreement with quantum mechanics;
still, the experiment has aroused some lively and controversial
debate (e.g., \cite{Sancho02,Peres02,BrEs05}). As pointed out by Short \cite{Short01},
Kim and Shih overlook the fact that the two width parameters in
question should be determined by the reduced quantum state of the
particle and thus should, according to quantum mechanics, satisfy
the uncertainty relation. Short gives an explanation of the
experimental outcome in terms of the imperfect imaging process which
leads to image blurring, showing that there is indeed no violation of the
uncertainty relation.

Finally, it seems that papers with claims of actual or proposed experiments indicating
violations of the uncertainty relation hardly ever pass the threshold of the refereeing
process in major journals. They appear occasionally as contributions to conference
proceedings dedicated to realistic (hidden variable) approaches to quantum mechanics.

\section{Conclusion}\label{conc}

In this exposition we have elucidated the positive role of the uncertainty principle as a necessary and sufficient condition for the possibility of approximately localizing position and momentum.  We have also
noted that approximate position measurements can allow a control of the disturbance of the
momentum. Uncertainty relations for position and momentum thus come in three variants: for the widths of probability distributions, for accuracies of joint measurements, and for the trade-off between the accuracy of a position measurement and the necessary momentum disturbance (and vice versa).

In his seminal paper of 1927, Heisenberg gave intuitive formulations of all three forms of uncertainty relations, but it was only the relation for state preparations that was made precise soon afterwards. It
took several decades until the conceptual tools required for a rigorous formulation of the two measurement-related uncertainty relations had become available. Here we identified the following
elements of such a rigorous formulation.

First, a theory of approximate joint  measurements of  position and momentum had to be developed; 
this possibility was opened up by the representation of observables as positive operator measures. Second, a criterion of what 
constitutes an approximate measurement of one observable by means of another must be based on 
operationally significant and experimentally relevant measures of inaccuracy or error. Here we 
discussed three candidate measures: standard error, a distance of observables, and 
error bars. For each of these, a universal Heisenberg uncertainty relation holds, showing that for any observable on phase space the marginal observables cannot both approximate position and momentum 
arbitrarily well.

The proofs of these uncertainty relations are first obtained for the distinguished class of 
covariant phase space observables, for which they follow mathematically from a form of
uncertainty relation for state preparations. This formal connection between the preparation and
joint measurement uncertainty relations is in accordance with a postulate formulated by 
N.~Bohr \cite{Bohr28} in his famous Como lecture of 1927 which states that the possibilities of 
measurement should not exceed the possibilities of preparation.
The uncertainty relation for a general approximate joint 
observable for position and momentum can then be deduced from that for some associated 
covariant joint observable. 

Apart from the limitations on the accuracy of joint approximations of position and momentum, 
we have found Heisenberg uncertainty relations which quantify the necessary
intrinsic unsharpness of two observables that are jointly measurable, \emph{provided} they 
are to be approximations of position and momentum, respectively. Both limitations are consequences
of the noncommutativity of position and momentum.

Finally, the idea of a measurement of (say) position disturbing the momentum has been made
precise by recognizing that a sequential measurement of measuring first position and then
momentum constitutes an instance of a joint measurement of some observables, of which
the first marginal is an (approximate) position and the second a distorted momentum observable.
The inaccuracy inherent in the second marginal gives a measure of the disturbance of momentum.
The joint measurement uncertainty relations can in this context be interpreted as a trade-off between
the accuracy of the first position measurement against the extent of the necessary disturbance of the
momentum due to this measurement.

Last we have surveyed the current status of experimental implementations of joint measurements and the question of experimental tests of the uncertainty principle. While there do not seem to exist any confirmed violations of the uncertainty principle, there do exist several experimental tests of uncertainty relations which have shown agreement with quantum mechanics.

\section*{Acknowledgements}

Part of this work was carried out during mutual visits by Paul Busch at the University
of Turku and by Pekka Lahti at Perimeter Institute. Hospitality and financial support by these host institutions are gratefully acknowledged. We are indebted to M.~Leifer and R.~Werner for helpful comments on an earlier draft version of this paper.

\section*{Appendix: Operations and instruments}
\addcontentsline{toc}{section}{Appendix: Operations and instruments}

In this appendix we recall briefly the concepts of an operation and an instrument, 
which are the basic  tools for describing the changes experienced by a quantum system
under the influence of a measurement or other interactions with external systems. 
In the Schr\"odinger representation these changes are described in terms
of the states of the system  whereas in the dual Heisenberg picture they are described in terms of the observables of the system. For more details, see e.g. \cite[Chapter 4]{QTOS}. 

Let $\trh$ be the Banach space of the trace class operators on a Hilbert space $\hi$ and let 
$\mathcal{L}(\trh)$ be the set of bounded linear mappings 
on $\trh$. We recall that a linear operator $\rho\in\trh$ is a state if it is positive, $\rho\geq 0$, and of trace one, $\tr{\rho}=1$.
A linear map $\phi:\trh\to\trh$ is an  {\em operation} if it is positive, that is, $\phi(\rho)\geq 0$ for all 
$\rho\geq 0$, and has the property
$0\leq \tr{\phi(\rho)}\leq 1$
for all  states $\rho$.
A positive linear map on $\trh$ is necessarily bounded, so that any operation $\phi$ is an element of  
$\mathcal{L}(\trh)$. 

An operation $\phi:\rho\mapsto\phi(\rho)$ comprises the description of the state change of a system under a measurement in the following way: if the initial state is $\rho$, the final state (modulo normalization) is given by $\phi(\rho)$ provided this is a nonzero operator. The number $\tr{(\phi(\rho)}$ gives the probability for the occurrence of the particular measurement outcome associated with $\phi$, and hence, for this particular state change.

The adjoint $\phi^*:\mathcal{L}(\hi)\to \mathcal{L}(\hi)$  of an operation $\phi:\trh\to\trh$, also called the {\em dual operation}, is defined by the formula
$\tr{\rho\phi^*(A)}=\tr{\phi(\rho)A}$, $A\in\mathcal{L}(\hi), \rho\in\trh$, and it is a normal positive linear map with the property
$0\leq\phi^*(I)\leq I$.

Using $\phi^*$, the probability for a measurement outcome associated with an operation $\phi$ can be expressed as $\tr{\phi(\rho)}=\tr{\rho\phi^*(I)}$ for all states $\rho$. Here the operator $\phi^*(I)$ is the effect representing the measurement outcome under consideration. This effect is uniquely determined by the operation $\phi$.

Let $\Omega$ be a nonempty set and $\mathcal A$ a $\sigma$-algebra of subsets of $\Omega$.
 An {\em instrument} (on the measurable space $(\Omega,\mathcal A)$) is a mapping $\stfm$ from the $\sigma$-algebra $\mathcal A$ to $\mathcal{L}(\trh)$ such that
\begin{enumerate}
\item[(i)] $\stfm(X)$ is an operation for all $X\in\mathcal A$;
\item[(ii)]   for each state  $\rho\in\trh$
the map $X\mapsto\tr{\stfm(X)(\rho)}$ is a probability measure.
\end{enumerate}
This means that an instrument is an operation valued measure.

An instrument $\stfm:\mathcal A\to\mathcal{L}(\trh)$ determines a unique observable $E:\mathcal A\to\lh$
by the condition
\begin{equation}
\tr{\rho E(X)}=\tr{\stfm(X)(\rho)},
\end{equation}
which is required to hold for all states $\rho\in\trh$ and for all  sets $X\in\mathcal A$. 

Let $\stfm$ be an instrument. The dual operations $\stfm(X)^\ast, X\in\mathcal A$, constitute 
the {\em dual instrument} $\stfm^\ast:\mathcal A\to\mathcal L(\hi)$ so that $\stfm^\ast(X)(A)= \stfm(X)^\ast(A)$ for any $X\in\mathcal A, A\in\lh$,  and thus
\begin{equation}
\tr{\rho\stfm^\ast(X)(A)}=\tr{A\stfm(X)(\rho)},
\end{equation}
for any  $X\in\mathcal A, A\in\lh, \rho\in\trh$.

In the application of this paper $(\Omega,\mathcal A)$ is the real Borel space
$(\R,\br)$.

\end{document}